\documentclass[preprint,authoryear,12pt]{elsarticle}

\usepackage{epsfig}

\usepackage{amssymb}

\usepackage[ps2pdf,%
a4paper=true,%
breaklinks=true,%
colorlinks=true,%
pdfauthor={First Author et al.},%
pdftitle={Template for manuscripts in Advances in Space Research}%
]{hyperref}

\newcommand{\be}{\begin{equation}}
\newcommand{\ee}{\end{equation}}
\newcommand{\bdm}{\begin{displaymath}}
\newcommand{\edm}{\end{displaymath}}

\def\degr{\hbox{$^\circ$}}
\def\sun{\hbox{$\odot$}}

\journal{Advances in Space Research}

\begin{document}

\begin{frontmatter}



\title{On the origin of the peculiar cataclysmic variable AE\,Aquarii}


\author{N.G.\,Beskrovnaya\corref{cor}}
\address{Central Astronomical Observatory of the RAS at Pulkovo, Pulkovskoe Shosse 65--1,
St.\,Petersburg, 196140 Russia} \cortext[cor]{Corresponding author}
\ead{beskrovnaya@yahoo.com}

\author{N.R.\,Ikhsanov\fnref{footnote1}}
\address{Central Astronomical Observatory of the RAS at Pulkovo, Pulkovskoe Shosse 65--1,
 St.\,Petersburg, 196140 Russia}
\fntext[footnote1]{Saint Petersburg State University, St.\,Petersburg, 198504 Russia}


\begin{abstract}

The nova-like variable AE\,Aquarii is a close binary system containing a red dwarf
and a magnetized white dwarf rotating with the period of 33\,seconds.
A short spin period of the white dwarf is caused by an intensive mass exchange between
the system components during a previous epoch. We show that a high rate of disk
accretion onto the white dwarf surface resulted in temporary screening of its magnetic
field and spin-up of the white dwarf to its present spin period. Transition of the white
dwarf to the ejector state occurred at a final stage of the spin-up epoch after its
magnetic field had  emerged from the accreted plasma due to diffusion. In the frame of
this scenario AE Aqr represents a missing link in the chain of Polars evolution and the
white dwarf resembles a recycled pulsar.

\end{abstract}


\end{frontmatter}

\parindent=0.5 cm

\section{Introduction}

AE~Aquarii is a peculiar nova-like star which exhibits rapid flaring
in almost all parts of the spectrum from radio to X-rays. It is a
non-eclipsing close binary system at a distance of $\sim 100\pm
30$\,pc with the orbital period $P_{\rm orb} \approx 9.88$\,h and
almost zero eccentricity. The system components are a K3--K5 red
dwarf and a white dwarf in a very unusual state. First, it rotates
with the period of $P_{\rm s} \approx 33$\,s. Second, it brakes so
rapidly that the spin-down power exceeds the bolometric luminosity
of the system. Finally, its magnetic field prevents the inflowing
material from approaching the star to a distance smaller than $(3-5)
\times 10^{10}$\,cm, which is a factor of 20 larger than the
corotation radius of the white dwarf and 40 times larger than the
radius of the white dwarf itself \citep[for detailed system
decription and corresponding refecences see
e.g.][]{Ikhsanov-Beskrovnaya-2012}.

These peculiar properties can be explained in terms of the
pulsar-like white dwarf scenario provided the surface magnetic field
of the white dwarf is in excess of 50\,MG \citep{Ikhsanov-1998,
Ikhsanov-etal-2004}. But how such a white dwarf was formed? We
suggest that the white dwarf in AE~Aqr was spun-up by intensive
accretion in a previous epoch. The high rate of disk accretion onto
the surface of the white dwarf resulted in temporary screening of
its magnetic field. This has allowed the white dwarf to reach its
present spin period. Transition of the white dwarf to the ejector
state occurred at a final stage of the spin-up epoch after its
magnetic field had emerged from the accreted plasma due to
diffusion. In the frame of this scenario AE~Aqr represents a missing
link in the chain of Polars evolution and the white dwarf resembles
a recycled pulsar.

\section{Accretion-driven spin-up epoch}

A hint to answer the question about the origin of the fast rotating
strongly magnetized white dwarf in AE Aqr is provided by a
discrepancy between the age of the white dwarf determined by its
cooling time and the spin-down time scale  $t_{\rm sd} \simeq P_{\rm
s}/2\dot{P_{\rm s}} \simeq 10^7$\,years \citep{de-Jager-etal-1994}. Indeed,
the age of a $M_{\rm wd} \sim 0.8\,M_{\sun}$ white dwarf with the
surface temperature $T_{\rm wd} \leq 16\,000$\,K is limited to $\geq
10^8$\,yr \citep{Schoenberner-etal-2000}. This exceeds the spin-down
timescale of the white dwarf in AE~Aqr by more than an order of
magnitude. Hence, the fast rotation of the white dwarf cannot be
connected with peculiarities of its origin but is a product of the
binary evolution which contained an epoch of rapid spin-up of the
degenerate component caused by intensive accretion onto its surface.

The accretion-driven spin-up of a white dwarf can be effective only
if the mass-transfer rate between the system components satisfies
the condition $\dot{M}_{\rm pe} > \dot{M}_{\rm crit}$. Here
$\dot{M}_{\rm crit} \simeq 10^{-7}\,{\rm M_{\sun}\,yr^{-1}}$ is a
critical value of the accretion rate at which the hydrogen burning
in the matter deposited onto the white dwarf surface is stable.
Otherwise, the spin behavior of the star will be similar to dwarf
novae in which spin-up of the degenerate component is prevented by
thermonuclear runaways followed by the expanded envelope mass-loss
spin-down of the white dwarf \citep{Livio-Pringle-1998}.

As shown by \citet{Meintjes-2002}, the mass transfer rate in AE~Aqr
during a previous epoch could be as high as $\dot{M}_{\rm pe} \sim
10^{19}-10^{20}\,{\rm g\,s^{-1}}$ ($\sim 10^{-7} - 10^{-6}\,{\rm
M_{\sun}\,yr^{-1}}$). If the magnetic field strength of the white
dwarf, and, correspondingly, its magnetospheric radius, remains
unchanged during this epoch, its spin period decrease down to 33\,s
on the time scale \citep{Ikhsanov-1999}
   \be\label{dur}
\Delta t_{\rm max} \geq \frac{2\pi I}{\dot{M}_{\rm pe}\
\sqrt{GM_{\rm wd} R_{\rm m}}}\ \left(\frac{1}{P_{\rm s}} - \frac{1}{P_{\rm
i}}\right) \simeq 2 \times 10^5\ I_{50} \dot{M}_{19}^{-1}
M_{0.8}^{-1/2} R_9^{-1/2} P_{33}^{-1}\ {\rm yr}.
    \ee
Here $I_{50}$ is the moment of inertia of the white dwarf in units
$10^{50}\,{\rm g\,cm^2}$ and $\dot{M}_{19} = \dot{M}/10^{19}\,{\rm
g\,s^{-1}}$. $M_{0.8}$, $R_9$ and $P_{33}$ are the mass, radius and
spin period of the white dwarf in units $0.8\,{\rm M_{\sun}}$,
$10^9$\,cm and $33$\,s, correspondingly. Finally, $P_{\rm i}$ is an
initial spin period of the white dwarf, which is assumed to satisfy
inequality $P_{\rm i} \gg 33$\,s.

The ultimate period which the white dwarf can reach in the process
of disk accretion is given by $P_{\rm min} = \max\{P_{\rm m}, P_{\rm
eq}\}$. Here $P_{\rm m}$ is a solution to equation  $R_{\rm m} =
R_{\rm cor}$, and $P_{\rm eq}$ is an equilibrium period defined by
equality of the spin-up torque, $K_{\rm su} = \dot{M}_{\rm pe}
\left(GM_{\rm wd} R_{\rm m}\right)^{1/2}$, and spin-down torque,
$K_{\rm sd} = (1/4) k_{\rm t} B_{\rm s}^2 R_{\rm wd}^6/R_{\rm
cor}^3$, applied to the white dwarf from the accretion flow. Here
$R_{\rm m}$ and $B_{\rm s}$ are the magnetospheric radius and the
magnetic field strength on the surface of the white dwarf at the
final stage of the accretion-driven spin-up, and  $k_{\rm t}$ is a
numerical coefficient. In the case of stationary accretion and under
the conditions of interest $P_{\rm eq} \leq P_{\rm m}$. Taking
$P_{\rm eq} = 33$\,s in equation $K_{\rm su} = K_{\rm sd}$ and
solving it for $B_{\rm s}$, we find that the observed spin period of
the white dwarf in AE~Aqr can be reached within the scenario of
accretion-induced spin-up provided $B_{\rm s} \leq B_0$, where
 \be
 B_0 \simeq 1.5\,{\rm MG}\ k_{0.3}^{-7/12} M_{0.8}^{5/6} R_{8.8}^{-3} P_{33}^{7/6} \dot{M}_{19}^{1/2}.
 \ee
and $k_{0.3} = k_{\rm t}/0.3$. This indicates that reconstructing
the evolutionary track of the system it is necessary to take into
account not only the spin evolution of the white dwarf \citep[as it
has been done by][]{Meintjes-2002}, but also the evolution of its
magnetic field.

 \section{Magnetic field screening during the spin-up epoch}

The magnetic field of the white dwarf may decrease during the
spin-up epoch due to screening by the accreting material
\citep{Bisnovatyi-Kogan-Komberg-1974}. The hypothesis about a
possibility to bury the magnetic field of accretors has been
actively investigated for neutron stars \citep{Konar-Choudhuri-2004,
Lovelace-etal-2005} and white dwarfs \citep{Cumming-2002}. The
efficiency of screening has been shown to depend on the mass
accretion rate and a duration of the intensive mass exchange between
the system components. Under favorable conditions the surface
magnetic field of a star can be reduced by a factor of 100.
Afterwards the field is expected to reemerge in the process of
diffusion through the layer of accreted plasma.

Following this hypothesis we can assume that prior to the epoch of
active mass exchange the magnetic field strength on the surface of
the white dwarf in AE~Aqr was close to its current value. At that
time the system was likely to behave as a Polar (since the
magnetospheric radius of the compact component under the condition
$\dot{M} \ll \dot{M}_{\rm pe}$ essentially exceeds its
circularization radius). The start of spin-up epoch was caused by
increase of the mass exchange rate up to $\dot{M}_{\rm pe} \geq
10^{19}\,{\rm g\,s^{-1}}$ due to red dwarf overfilling its Roche
lobe. This resulted in decrease of the magnetospheric radius of the
white dwarf down to $R_{\rm m}^{\rm (i)} \leq 10^{10}$\,cm and
subsequent formation of the accretion disk in the system. The
accretion of matter onto the surface of the white dwarf in this case
could occur under condition $R_{\rm m}^{\rm (i)} < R_{\rm cor}$
which was satisfied provided the initial spin period of the white
dwarf was $P_{\rm i} \geq 11$\,min.

The field of the compact object was found to be strongly screened by
plasma accumulating in its polar caps  for accretion rates greater
than the critical value $\geq 3 \times 10^{16}\,{\rm g\,s^{-1}}$
\citep{Cumming-2002}. Because of surface field decay the
magnetospheric radius of the white dwarf is decreasing and,
correspondingly, the area of the hot spots on its surface is
increasing. The maximum possible factor of field reduction during
the epoch of intensive accretion is limited to  $\sim
\left(1/\sin{\theta_{\rm i}}\right)^{7/2} \sim 125$, where
$\theta_{\rm i} = \arcsin\left(R_{\rm wd}/R_{\rm
m}^{(i)}\right)^{1/2}$ is the opening angle of the accretion column
at the beginning of spin-up epoch. This implies that at the final stages of
spin-up epoch the magnetic field of the white dwarf did not exceed
1\,MG and, hence, could not prevent decrease of the spin period down
to its current value.

The spin-up time of the white dwarf  with account for screening of
its magnetic field in the process of accretion can be evaluated by
solving the equation $I \dot{\omega}_{\rm s} = \dot{M}_{\rm pe}
\left(GM_{\rm wd} R_{\rm cor}\right)^{1/2}$
\citep{de-Jager-etal-1994} based on the assumption that the
magnetospheric radius of the white dwarf is decreasing at the same
rate that its corotation radius. The solution to this equation
  \be
\Delta t_{\rm min} = \frac{3}{4} \frac{(2 \pi)^{4/3} I}{\dot{M}_{\rm
pe} (GM_{\rm wd})^{2/3} P_{\rm s}^{4/3}},
   \ee
determines the minimum duration of the spin-up epoch. The amount of
matter accumulated on the white dwarf surface during this period can
be estimated as \be\label{deltama} \Delta M_{\rm a} = \dot{M}_{\rm
pe} \Delta t_{\rm min}  =  \frac{3}{4} \frac{(2 \pi)^{4/3}
I}{(GM_{\rm wd})^{2/3} P_{\rm s}^{4/3}}.
  \ee
After the accretion epoch is over the surface magnetic field of the
white dwarf is gradually increasing due to diffusion of the buried
field through the layer of screening plasma. The diffusion timescale
of the field can be estimated as $t_{\rm diff} \sim 4 \pi \sigma
h^2/c^2$, where $\sigma$ is the electron conductivity and $h = p/\rho g$
is the pressure scale height. Here $p \sim 6.8 \times 10^{20}
\rho_5^{5/3}\,{\rm erg\,cm^{-3}}$ is the pressure of
non-relativistic degenerate gas, $\rho$ in the plasma density at the
base of the screening layer  ($\rho_5 = \rho/10^5\,{\rm
g\,cm^{-3}}$) and $g = GM_{\rm wd}/R_{\rm wd}^2$. \citet{Cumming-2002} has shown
that reemergence of the field of the white dwarf having undergone the stage of active accretion occurs on the timescale
 \be
\tau_{\rm diff} \simeq 3 \times 10^8 \left(\Delta M_{\rm
a}/0.1\,{\rm M_{\sun}}\right)^{7/5}\,{\rm yr}.
 \ee

 \section{The pulsar-like white dwarf formation}

An appearance of a rapidly rotating  highly magnetic white dwarf can
be expected only under the condition  $\tau_{\rm diff} \leq t_{\rm
sd}$. Otherwise the spin period of the compact component will
essentially increase on the timescale of field reemergence. Solving
this inequality for the parameters of AE~Aqr we find
 \be
\Delta M_{\rm a} \leq 0.009\ P_{33}^{5/7} \left(\frac{\dot{P_{\rm s}}}{5.64
\times 10^{-14}\,{\rm s\,s^{-1}}}\right)^{-5/7}\ {\rm M_{\sun}}.
 \ee
Putting this value to Eq.~(\ref{deltama}) leads to a conclusion that
the origin of an ejecting white dwarf in AE~Aqr can be explained in
terms of accretion-induced spin-up provided its moment of inertia is
 \be
 I \leq 6 \times 10^{49}\ P_{33}^{4/3} \left(\frac{M_{\rm wd}}{{\rm M_{\sun}}}\right)^{2/3}
 \left(\frac{\Delta M_{\rm a}}{0.009\,{\rm M_{\sun}}}\right)\ {\rm g\,cm^2}.
 \ee
According to \citet{Andronov-Yavorskij-1990}, this condition is
satisfied for white dwarfs with the mass in the range $1.1-1.2\,{\rm
M_{\sun}}$.

The result obtained allows  to make some conclusions about the
system parameters in general. First of all, relatively large mass of
the white dwarf indicates that the angle of orbital inclination is
close to  $50^{\degr}$. This value is within the range of permitted
values for this parameter \citep{Welsh-etal-1995}. It implies the
mass of the red dwarf companion in excess of $0.7\,{\rm M_{\sun}}$
and, accounting for its tidal distortion
\citep{van-Paradijs-etal-1989}, lead to the estimate of its tidal
radius (along the system major axis) comparable to the radius of its
Roche lobe. Finally, a
correction of the inclination angle (its shift towards lower values)
leads us to conclusion that the velocity of the gaseous stream in
the Roche lobe of the white dwarf is somewhat greater than initially
adopted and, hence, the distance of the stream closest approach  to
the white dwarf is somewhat less than previously estimated. This
fact has to be taken into account in the modeling of the mass
transfer in the system in the present epoch.

  \section{Conclusions}

Our analysis shows that the origin of the peculiar white dwarf in
AE~Aqr can be connected with intensive mass exchange between the
system components in a previous epoch. In the process of accretion
which took place in that epoch, the material deposited from the
accretion disk onto the white dwarf surface temporarily screened the
internal magnetic field of the white dwarf thus making possible
accretion-induced spin-up up to its current level. The transition of
the white dwarf into the ejector state was caused by reemerging of
the magnetic field by diffusion through the layer of accreted
matter.

Relatively large age of the white dwarf  ($\sim 10^9$\,yr) derived
from its average surface temperature, limitation on its intrinsic
spin period ($P_{\rm i} > 11$\,min) and our estimate of its dipole
magnetic moment ($\mu \sim 10^{34}\,{\rm G\,cm^3}$) make us to
suggest that before the spin-up epoch AE~Aqr could manifest itself
as a Polar. During the spin-up epoch its X-ray luminosity exceeded
$10^{36}\,{\rm erg\,s^{-1}}$ and the system could be seen as
extremely bright Intermediate Polar. One cannot exclude that during
the final phase of spin-up, the accretion of matter onto the white
dwarf surface occurred directly from the accretion disk (as in
non-magnetic CVs) and a  component pulsing at the spin period of the
white dwarf was not present in the X-ray emission from the system.
The duration of the present epoch is likely to be determined by the
spin-down time-scale of the white dwarf which is close 10\,million
years. At the end of this epoch one can expect dissipation of
electric currents in the white dwarf magnetosphere and its
transition to the propeller state. Further the system will appear as
a Polar.

In the frame of this scenario AE~Aqr can be considered as a missing
evolutionary link in the evolution of Polars, with its origin
resembling in some aspects evolutionary scenario for recycled
pulsars. At the same time, the analogy with  evolution of recycled
pulsars is incomplete since before the spin-up epoch the white dwarf
was in the accretor state with relatively slow rotation. Thus, in
the case of AE~Aqr we deal with essentially new evolutionary stage
of low-mass binaries requiring introduction of a new subclass of
cataclysmic variables. The degenerate objects in the systems from
this subclass are in the ejector state. Intensive matter outflow
from a system and a presence of high-luminous non-thermal component
in its emission can be considered as indirect attribute of these
systems, while the contribution of accretion luminosity to their energy budget
is insignificant.

\section*{Acknowledgements}
This work was partly supported by the Russian Foundation of Basic Research
under grant N\,13-02-00077 and the Program of Presidium of Russian
Academy of Sciences N\,21.


\end{document}